\documentclass[twocolumn]{aastex62}
\pdfoutput=1 
\usepackage{amsmath,amstext}
\usepackage[T1]{fontenc}
\usepackage{apjfonts} 
\usepackage[figure,figure*]{hypcap}
\usepackage{graphicx}

\shorttitle{LMC Chemical Enrichment History}
\shortauthors{Chilingarian \& Asa'd}

\begin{document}

\title{Using Star Clusters as Tracers of Star Formation and Chemical Evolution: the Chemical Enrichment History of the Large Magellanic Cloud\footnote{All star cluster spectra presented in this work and their best-fitting stellar population templates are available at the following permanent address: \url{https://doi.org/10.5281/zenodo.1203653}}}

\author[0000-0002-7924-3253]{Igor V. Chilingarian}
\altaffiliation{e-mail: igor.chilingarian@cfa.harvard.edu}
\affiliation{Smithsonian Astrophysical Observatory, 60 Garden St. MS09, Cambridge, MA, 02138, USA}
\affiliation{Sternberg Astronomical Institute, M.V. Lomonosov Moscow State University, 13 Universitetsky prospect, Moscow, 119991, Russia}
\author{Randa Asa'd}
\altaffiliation{Summer Visiting Scientist at Harvard-Smithsonian Center for Astrophysics}
\affiliation{Physics Department, American University of Sharjah, P.O. Box 26666, Sharjah, UAE}

\begin{abstract}
The star formation (SFH) and chemical enrichment (CEH) histories of Local Group galaxies are traditionally studied by analyzing their resolved stellar populations in a form of color-magnitude diagrams obtained with the Hubble Space Telescope. Star clusters can be studied in integrated light using ground-based telescopes to much larger distances. They represent  snapshots of chemical evolution of their host galaxy at different ages. Here we present a simple theoretical framework for the chemical evolution based on the instantaneous recycling approximation (IRA) model. We infer a CEH from a SFH and vice versa using observational data. We also present a more advanced model for the evolution of individual chemical elements which takes into account the contribution of supernovae type Ia. We demonstrate that ages, iron and $\alpha$-element abundances of 15 star clusters derived from  fitting of their integrated optical spectra reliably trace the CEH of the Large Magellanic Cloud obtained from resolved stellar populations in the age range 40~Myr$<t<$3.5~Gyr. The CEH predicted by our model from the global SFH of the LMC agrees remarkably well with the observed cluster age--metallicity relation. Moreover, the present day total gas mass of the LMC estimated by the IRA model ($6.2\cdot10^8 M_{\odot}$) matches within uncertainties the observed H{\sc i} mass corrected for the presence of molecular gas ($5.8\pm0.5\cdot10^8 M_{\odot}$). We briefly discuss how our approach can be used to study SFHs of galaxies as distant as 10~Mpc at the level of detail that is currently available only in a handful of nearby Milky Way satellites.
\end{abstract}


\section{Introduction and Motivation}

Strong starburst events or periods of quiescent evolution that take place during the lifetime of a galaxy form consequent generations of stars enriching the interstellar medium at the end of their lives with metals produced inside them. Gas-rich mergers and gas infall from companions or cosmic filaments supply additional fuel for star formation, while the feedback from supernovae and stellar winds slow it down and can eventually quench it completely \citep{Dekel86,Chiosi02,Matteucci94,Bekki05}. Therefore, stellar populations of a galaxy contain a fossil record of its evolutionary path. The star formation and chemical enrichment histories of a galaxy are most directly revealed through analysis of its individual stars and resolved star clusters \citep[see e.g.][and references therein]{Dirsch2000, Carrera08, Maschberger11, Livanou13, Piatti17}. However, this approach only applies to galaxies at distances up-to $\sim$1~Mpc that are close enough to be resolved. The reconstruction of star formation and chemical enrichment histories of a galaxy from the integrated light usually referred to as ``unresolved stellar populations analysis'', is an ill-conditioned inverse problem which is subject to degeneracies between parameters and often yields non-unique solutions. 

Here we propose to use massive star clusters as direct tracers of galaxy star formation and chemical enrichment histories. Massive star clusters ($M>10^4 M_{\odot}$) are luminous enough so that their integrated light spectra can be collected out to much larger distances ($\sim10$~Mpc) than the limit of resolved stellar population studies \citep{Asad14,Asad16P,Asad16}. By using multi-object spectroscopy, one can collect up-to a few hundreds of star cluster spectra in a nearby galaxy at once. On the other hand, stars in such clusters are formed almost instantaneously from the available material, so that the effects of self-enrichment of the interstellar medium (ISM) with metals can be neglected. Hence, a star cluster can be considered as a snapshot of the chemical enrichment history of a galaxy region where it was formed at the time of its formation. This also means, that star cluster spectra should be well described by simple stellar population (SSP) models (i.e. models which are constructed of stars of the same age and metallicity). 

In this work we demonstrate that we can reproduce the chemical enrichment history of the Large Magellanic Cloud (LMC) using integrated spectra of star clusters. We chose LMC for several reasons. It is superior to our Galaxy for such studies because it is relatively small and the dynamical processes in its disk (that mix stellar populations, such as radial migration) are much less important than in giant disks \citep[see e.g.][]{Schroyen13}. LMC is a low-mass galaxy located in a group dominated by two giants (Milky Way and Andromeda), therefore, it must have experienced a very few (if any) minor mergers with gas-rich satellites that brought metal-poor gas into it and affected the chemical enrichment history. LMC has very shallow if any radial metallicity gradient \citep{2010MNRAS.408L..76F}, hence, the exact location of a particular star cluster in its disk is not important for the analysis of the global properties. Moreover, the SFH in the LMC is tightly connected to the galaxy evolution in the Local Group and the mass assembly history of the Milky Way. Hence, analyzing the LMC is an important step to better understanding of the evolution of the Local Group. Finally, the LMC is at the right distance (51~kpc) to allow both resolved and unresolved stellar population analysis of its star clusters \citep{Asad12,Asad13}, and the foreground dust extinction towards LMC clusters is much better constrained than in regions close to the Galactic plane. This allows us to compare our data analysis for integrated spectra against resolved stellar populations at young and intermediate ages of stellar populations.

The paper is organized as follows: in \emph{Section~2} we present some theoretical aspects of the chemical evolution and explain how they can be applied to observational data; \emph{Section~3} describes our spectroscopic observations and data reduction; in \emph{Section~4} we provide the results of stellar population analysis for 15 LMC star clusters; and in \emph{Section~5} we connect observations with the theory and discuss further prospective of using star clusters to study chemical evolution of galaxies.

\section{Connecting star formation history to chemical enrichment}

In order to link the star formation rate (SFR) $\psi(t)$ with the metal content of the ISM $Z(t)$, one needs to develop a model of the galactic chemical evolution. The simplest model called the ``instantaneous recycling approximation'' \citep[IRA, see e.g.][]{Schmidt63,Pagel97,Matteucci16} assumes that all stars less massive than 1~M$_{\odot}$ live forever and stars above this mass die instantaneously enriching the interstellar medium with heavy elements formed in their interiors.

The main drawback of this approach is that it does not account for delayed chemical enrichment by asymptotic giant branch (AGB) stars and type Ia supernovae, which enter into action hundreds of millions to billions of years after the beginning of a star formation episode \citep[see e.g.][and references therein]{NKT13}. Therefore, in order to describe chemical evolution of species produced by different astrophysical phenomena, we need to expand the IRA model and include the contributions by Type Ia SNe and AGB stars. In this work we analyze absorption line spectra of stellar populations, which allow us to assess iron and $\alpha$-element abundances but do not provide accurate estimates for oxygen and carbon. Because AGB stars contribute neither to iron nor to $\alpha$-element enrichment, we restrict our analytic model to IRA$+$SNIa.

\subsection{Instantaneous recycling approximation}

The IRA model of chemical evolution \citep[see][for detailed description]{Spitoni17} is defined by the following three parameters: $R$ -- the returned mass fraction of gas that goes back to ISM after stars evolve and can be recycled for star formation, $y_Z$ -- the metal yield per stellar generation that is a mass fraction of heavy elements formed in stars and returned into the ISM, and $\lambda$ -- the outflow (or galactic wind) coefficient \citep{Matteucci83,Matteucci12} in a simple model of galactic winds directly proportional to the SFR: $W(t)=\lambda \psi(t)$. The two former parameters $R$ and $y_Z$ depend on the metallicity and the initial mass function (IMF), however, the metallicity effects turn out to be minor compared to those of the IMF \citep{Vincenzo16}. \citet{Spitoni17} computed those coefficients and obtained the following values: $R=0.287$, $y_Z=0.0301$ for the \citet{Salpeter55}  and \citet{Kroupa01} IMFs and $R=0.441$, $y_Z=0.0631$ for the \citet{Chabrier03} IMF. We do not consider gas infall because: (i) the infall is usually considered as a process exponentially declining in time \citep{Chiosi80} but our study is concentrated on the last 3--4~Gyr of the LMC history; (ii) in the case of LMC located in the vicinity of the much more massive Milky Way, the infalling gas would likely be accreted by the Milky Way.

In this framework, the chemical evolution is described by the following system of differential equations (see eq.~7 in \citealp{Spitoni17} without the infall parts) for the gas mass $M_\mathrm{gas}(t)$ and the total mass of metals in the ISM $M_Z(t)$:
\begin{equation}
\small{
\begin{cases}
\dot{M}_\mathrm{gas}(t) =  -\big(1 - R \big) \psi(t) - \lambda \psi(t)  & \\
\dot{M}_Z(t) = \big(-Z(t)+y_Z \big)\big(1-R\big) \psi(t) -\lambda Z(t) \psi(t) & \\
M_Z(t) = M_\mathrm{gas}(t) Z(t)
\label{eq_sys_chemevo}
\end{cases} }
\end{equation}

We solve it for the two cases: (i) when the star formation history $\psi(t)$ is an observable quantity and we would like to predict the chemical evolution caused by self-enrichment $Z(t)$; and (ii) when we observe the chemical evolution $Z(t)$ and try to recover the star formation history $\psi(t)$.

The equation for $M_\mathrm{gas}$ can be integrated as:
\begin{equation}
M_\mathrm{gas}(t) = M_\mathrm{gas}(0) - (1-R+\lambda) \int_0^t \psi(T) dT
\label{eq_mgas}
\end{equation}

\noindent Then the equation for the chemical enrichment becomes:
\begin{equation}
\dot{Z}(t) = y_Z (1-R) \frac{\psi(t)}{M_\mathrm{gas}(0) - (1-R+\lambda) \int_0^t \psi(T) dT},
\label{eq_zevo}
\end{equation}

\noindent and we can then derive the chemical enrichment history as:
\begin{align}
Z(t) = Z(0) + y_Z (1-R) \int_0^t \frac{\psi(T)}{M_\mathrm{gas}(T)} dT = \nonumber \\
Z(0) + \frac{y_Z(1-R)}{1-R+\lambda} \mathrm{ln}\frac{M_\mathrm{gas}(0)}{M_\mathrm{gas}(0) - (1-R+\lambda) \int_0^t \psi(T) dT}
\label{eq_zevo_solved}
\end{align}

\noindent If we derive $\psi(t)$ from observations (e.g. color-magnitude diagrams or CMDs), we can then predict the metal content of the ISM by integrating Eq.~\ref{eq_zevo_solved} numerically. Given that $\psi(t)$ is non-negative, the integral in the denominator will monotonically grow, so is the metallicity. This solution has three free parameters, the initial ISM metallicity $Z(0)$ and gas mass $M_\mathrm{gas}(0)$, and the outflow coefficient $\lambda$. It is important to point out that $Z(0)$ and $M_\mathrm{gas}(0)$ correspond to the start of time period covered by observations that does not have to be the same as the actual galaxy formation epoch. Modern CMD analysis usually provides quite accurate metallicity estimates for old stellar populations which can be used as $Z(0)$. At the same time, $M_\mathrm{gas}(0)$ and $\lambda$ are degenerated. However, if we consider several regions in a galaxy where we expect similar global conditions, we would expect $\lambda$ to stay the same across all of them.

If we observe the chemical enrichment e.g. traced by star clusters or associations and we aim at recovering the SFH, we will have to solve the integral Equation~\ref{eq_zevo_solved} for $\psi(t)$ by isolating the integral on one side and taking a derivative on $t$:
\begin{equation}
\psi(t)=\frac{\dot{Z}(t) M_\mathrm{gas}(0)}{y_Z(1-R)}\exp(-\frac{1-R+\lambda}{y_Z(1-R)}(Z(t)-Z(0)))
\label{eq_psisol}
\end{equation}
\noindent Because this expression includes $\dot{Z}(t)$ that is subject to noise in the data,  metallicity measurements should properly sample the range of ages where the SFH needs to be recovered. We also stress that $\dot{Z}(t)$ should stay positive, which reflects the assumption that our model does not include the infall of metal poor gas and, therefore, the ISM metallicity can only increase in time.

\subsection{Chemical evolution model including the contribution of Type Ia SNe}

The importance of type Ia supernova for the iron enrichment was for the first time addressed by \citet{Tinsley80} and then studied observationally and theoretically by \citet{GR83,NTY84,MG86}. Massive stars exploding as core collapse supernova (SN~Ibc/SN~II) enrich the ISM predominantly with $\alpha$-elements \citep{WW95,NKT13}. Because massive stars ($M>8M_{\odot}$) are short-living ($10^6-10^7$~yr), the IRA model is fully applicable to the abundance prediction of oxygen and $\alpha$-elements.

On the other hand, Type Ia SNe produce substantial amounts of iron peak elements  \citep{NTY84,Iwamoto+99,NKT13} but the timing of the enrichment is very different. Type Ia SNe originate from either white dwarfs exceeding the Chandrasekhar mass limit because of the gas accretion from a companion star \citep{WI73,KLMT93}, a so-called single degenerate or (SD); or from two white dwarfs merging because their orbit shrinks as the orbital energy is consumed by the emission of gravitational waves \citep{IT84,IT85}, a so-called double degenerate (DD) scenario. In both scenarios it takes significant time before the first Type Ia SNe go off, therefore iron enrichment lags that of oxygen and $\alpha$-elements \citep{Matteucci94}.

The approach we follow here treats the chemical evolution using a two-component model: the IRA plus the contribution of Type Ia SNe. The idea is described in detail in \citet{VMS17}, but here we introduce some modifications: instead of using the local Schmidt--Kennicutt star formation law and dealing with the ISM surface density in a galactic disk, in case of LMC we can instead use the global galaxy-wide SFH because the gas is relatively well mixed and no significant metallicity gradient is observed in the LMC disk.

Given the stellar initial mass function, the fraction of close binary stars and the distribution of their orbital separation, one can compute the delay time distribution (DTD) of SN~Ia explosions for different progenitor models \citep[see][and references therein for a detailed description]{MSRV09}. Then, the SN~Ia rate $R_{\mathrm{Ia}}(t)$ as a function of time can be computed as a convolution of the DTD with the SFH $\psi(t)$:

\begin{equation}
R_{\mathrm{Ia}}(t) = C_{Ia} \int_0^t DTD(t) \psi(t-T) dT.
\label{eq_ria}
\end{equation}

\noindent Here $C_{Ia}$ is a normalization coefficient selected to match the observed SN~Ia rate, 2 per 1,000~$M_{\odot}$ formed in stars \citep{VMS17}. Then we can rewrite Eq.~\ref{eq_sys_chemevo} for abundances of individual elements $X(t)$ with the corresponding total masses $M_X(t)$ and include the SN~Ia contribution into it:
\begin{equation}
\small{
\begin{cases}
\dot{M}_\mathrm{gas}(t) =  -\big(1 - R + \lambda \big) \psi(t) + \langle m_{\mathrm{all}} \rangle R_{\mathrm{Ia}}(t) & \\
\dot{M}_X(t) = -X(t)\big(1-R+\lambda \big) \psi(t) + y_X \big(1-R\big)\psi(t) +\langle m_{X} \rangle R_{\mathrm{Ia}}(t) & \\
M_X(t) = M_\mathrm{gas}(t) X(t) & \\
\cdots
\label{eq_sys_chemevo_full}
\end{cases} }
\end{equation}
\noindent Here the notations for the IRA part are similar to those in Eq.~\ref{eq_sys_chemevo}, $R_{\mathrm{Ia}}(t)$ is the SN~Ia rate, $\langle m_{X} \rangle$ is the yield of the element $X$ from SN~Ia explosions, and $\langle m_{\mathrm{all}} \rangle$ is the combined yield of all chemical species from Type Ia SNe, which is very close to 1.343~$M_{\odot}$ for all SD and DD models presented in \citet{Iwamoto+99}.

The equation for $M_\mathrm{gas}$ can be integrated in a similar way to the IRA model as:
\begin{equation}
M_\mathrm{gas}(t) = M_\mathrm{gas}(0) - (1-R+\lambda) \int_0^t \psi(T) dT + \langle m_{\mathrm{all}} \rangle \int_0^t R_{\mathrm{Ia}}(T) dT
\label{eq_mgas_full}
\end{equation}
\noindent Here we notice an important difference compared to the IRA solution: Type Ia SNe create a positive contribution to the total gas mass, therefore it does not have to monotonically decrease over time as it does in the IRA framework and the exact behavior of $M_\mathrm{gas}(t)$ depends on the SFH.

Then the equation for the chemical enrichment becomes:
\begin{equation}
\dot{X}(t) = \frac{R_{\mathrm{Ia}}(t)(\langle m_{X} \rangle - \langle m_{\mathrm{all}} \rangle X(t))}{M_\mathrm{gas}(t)} +\frac{y_X (1-R) \psi(t)}{M_\mathrm{gas}(t)},
\label{eq_zevo_full}
\end{equation}
\noindent It can no longer be integrated analytically and requires numerical methods to be used. The free parameters defining the initial conditions are similar to the IRA, the starting gas mass $M_\mathrm{gas}(0)$ and the initial individual abundances of every element $X(0)$ instead of the initial average metallicity $Z(0)$. The integral equation to recover $\psi(t)$ from observed $X(t)$ also becomes very complex and requires inversion techniques with regularization to be used.

We solve Eq.~\ref{eq_mgas_full} for the IRA return fractions and net yields per generation of silicon and iron computed by \citet{VMS17} for the Kroupa IMF: $R=0.285; y_{Si}=8.5\cdot 10^{-4}; y_{Fe}=5.6\cdot 10^{-4}$. For the Type Ia SNe treatment we use the DTD for SD progenitors by \citet{MR01} and yields for the \emph{W7} model from \citet{Iwamoto+99}. Our code has also an option to use the mixed DD/SD DTD by \citet{MDVP06} and the yields for \emph{WDD} and \emph{CDD} models. We chose silicon and not magnesium because the $y_{Si}$ were already computed in \citet{VMS17} and the models of stellar populations, which we use for the analysis of star cluster spectra have the same abundance scaling for all $\alpha$-elements.

\section{Spectroscopic observations and data reduction}

\begin{figure}
\includegraphics[width=\hsize]{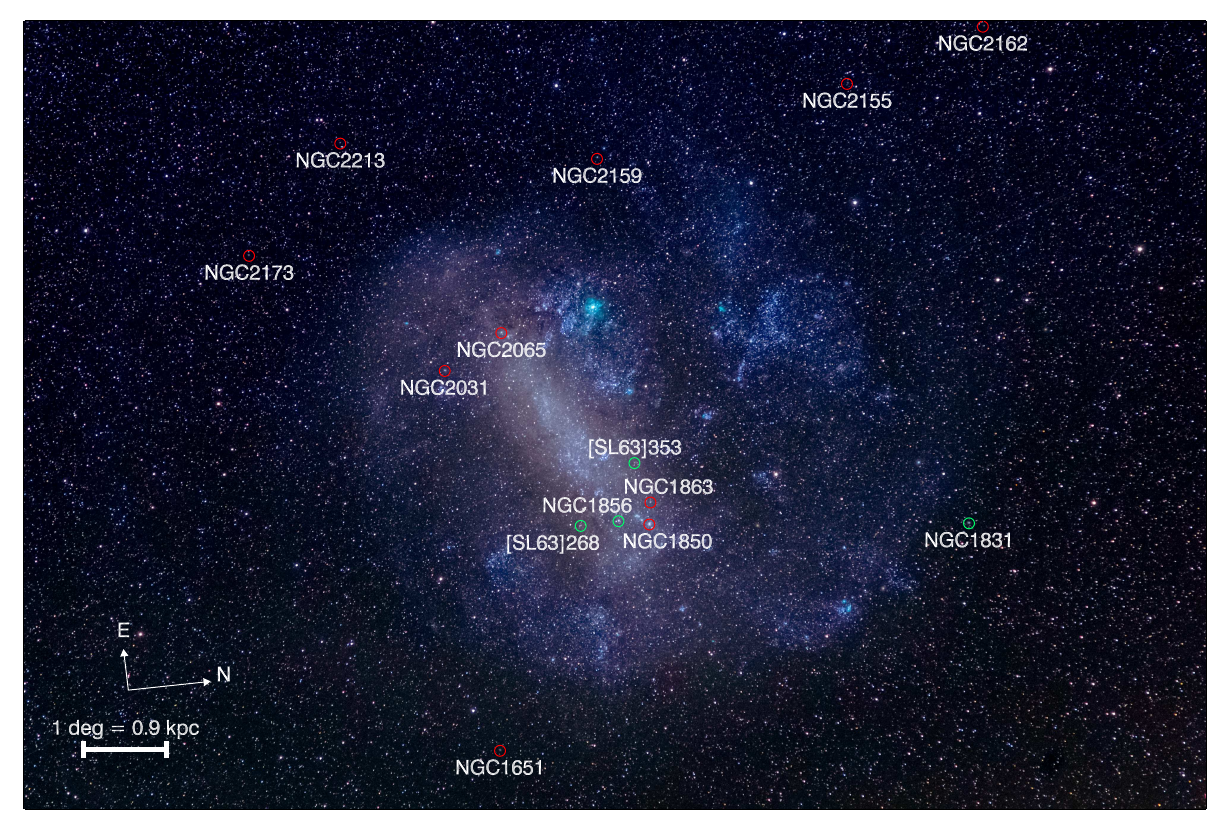}
\caption{Location of star clusters from our sample (green and red for Magellan and SOAR data respectively) in the Large Magellanic Cloud. \object{NGC~2249} is located slightly outside the top edge of the image. The underlying image was obtained by IC and R.~Mu\~noz in Feb/2016 using an amateur digital camera.\label{figlmcmap}}
\end{figure}

Our spectroscopic sample consists of two datasets (see Fig.~\ref{figlmcmap} showing the distribution of our observed clusters spanning a large region of the LMC). The first one comprises 11 clusters observed with the Goodman Spectrograph \citep{2004SPIE.5492..331C} on the 4-m SOAR telescope in the long-slit low resolution mode (1.03~arcsec wide slit; 600~gpm VPH grating; R=1500; $3600<\lambda<6250$~\AA) in December 2011 \citep{Asad13}. The second dataset includes 4 clusters observed with the 6.5-m Magellan Baade telescope in November 2016 using the intermediate resolution  Magellan Echelle (MagE, \citealp{2008SPIE.7014E..54M}) spectrograph (0.5~arcsec wide slit; R=7000; $3300<\lambda<9500$~\AA). Our sample was selected from clusters which had literature data on their ages derived from the analysis of CMDs obtained mostly with the Hubble Space Telescope. We selected the objects, which sampled the age range from 10~Myr to 3~Gyr.

In order to collect an integrated spectrum, we scanned a cluster across the slit using the non-sideral tracking. All star clusters from our sample have sizes between 0.8 and 3~arcmin, which provided enough area along the Goodman spectrograph slit to perform sky subtraction. MagE slit is only 10-arcsec long, therefore we had to (i) develop a script using the telescope control software, which scans a desired square region of the sky, e.g. 30$\times$30~arcsec through the slit in several back-and-forth passes during an exposure time set as a parameter (e.g. 900~sec); (ii) take offset sky exposures typically 2--3~arcmin away from the cluster. Not only our approach allowed us to obtain integrated spectra of extended objects, but we also sampled background close enough to each cluster in order to remove the contribution from the LMC disk.

We reduced our SOAR long-slit data using our generic long-slit / IFU data reduction package implemented in IDL. The data reduction procedure was almost identical to that for Gemini GMOS long-slit spectra described in detail in \citet{Francis12}: we started with bias subtraction and flat fielding, then built a two-dimensional wavelength solution along the slit and used it to construct an oversampled sky model using the algorithm proposed by \citet{Kelson01}, then subtracted the sky background and performed relative flux calibration using a spectrum of a spectrophotometric standard star. At the end, we integrated the spectrum along the slit and created the final one-dimensional data product. The error frames were created using the photon statistics and processed through the same steps in order to obtain flux uncertainties.

For MagE data we developed a dedicated data reduction pipeline, which is too complex to describe it here in detail\footnote{The beta-version of the MagE pipeline is available for testing purposes here: \url{https://bitbucket.org/chil_sai/mage-pipeline}}. The pipeline merges Echelle orders and produces a sky subtracted flux calibrated two-dimensional spectrum corrected for telluric absorption and a corresponding flux uncertainty frame. We then collapse the spectrum along the slit and obtain a one-dimensional flux calibrated data product.
  
\section{Star cluster ages and metal abundances from integrated light spectra}

\subsection{Data analysis technique: ages and [Z/H]}

We used the {\sc nbursts} full spectrum fitting technique \citep{Chilingarian07,Chilingarian07b} to obtain ages and average metallicities [Z/H] of star clusters in our spectroscopic sample. We fitted each spectrum against a grid of SSP models and obtained the best-fitting SSP age and metallicity as well as a radial velocity of each cluster using a nonlinear minimization. We did not fit the internal velocity dispersion of star clusters because it stayed much below the limits imposed by the spectral resolution of our data. The fitting algorithm includes a multiplicative polynomial continuum, which makes it insensitive to the dust extinction towards the cluster and to possible flux calibration imperfections in the data. We used version~11 of a low resolution ($R=2300$) SSP model grid \citep{Vazdekis10} with Solar-scaled abundance ratios covering the wavelength range from 3600 to 7400~\AA\ based on the MILES stellar library \citep{SanchezBlazquez06} computed using BaSTI isochrones \citep{Bertelli94,2004ApJ...612..168P} for the Kroupa IMF. 

In order to analyze SOAR spectra, which have lower spectral resolution than SSP models, we convolved the SSP grid with the wavelength-dependent SOAR spectral line spread function (LSF) derived from the fitting of a twilight spectrum against a high-resolution Solar spectrum. We quadratically subtracted the intrinsic LSF of MILES models from the SOAR LSF, which we derived. Because the MagE spectral resolution is higher than that of MILES spectra, we convolved the observed spectra to match the spectral LSF of MILES similarly to the analysis of high-resolution VLT spectra presented in \citet{Chilingarian11}.

\begin{deluxetable}{lrclrcll}
\tablecolumns{8}
\tablecaption{Ages and metallicities of LMC cluster obtained with the {\sc nbursts} full spectrum fitting \label{tablmcclstpop}}
\tablehead{
\colhead{Object} & \multicolumn{3}{c}{$t$, Myr} &
\multicolumn{3}{c}{[Fe/H], dex} & \colhead{Dataset}
}
\startdata
\object{NGC1651} & 1353 &$\pm$& 70 & -0.25 &$\pm$& 0.03& SOAR \\
\object{NGC1831} &  393 &$\pm$&  20 & -0.14 &$\pm$& 0.02& Magellan \\
\object{NGC1850} &   42 &$\pm$&   2 & -0.12 &$\pm$& 0.01& SOAR \\
\object{NGC1856} &  320 &$\pm$&  16 & -0.13 &$\pm$& 0.01& Magellan \\
\object{NGC1863} &   47 &$\pm$&   3 & -0.18 &$\pm$& 0.01& SOAR \\
\object{NGC2031} &  140 &$\pm$&   7 & -0.14 &$\pm$& 0.02& SOAR \\
\object{NGC2065} &  124 &$\pm$&   6 & -0.16 &$\pm$& 0.02& SOAR \\
\object{NGC2155} & 3757 &$\pm$& 190 & -0.64 &$\pm$& 0.03& SOAR \\
\object{NGC2159} &  121 &$\pm$&   6 & -0.13 &$\pm$& 0.03& SOAR \\
\object{NGC2162} & 1298 &$\pm$&  60 & -0.32 &$\pm$& 0.04& SOAR \\
\object{NGC2173} & 1423 &$\pm$&  70 & -0.37 &$\pm$& 0.02& SOAR \\
\object{NGC2213} & 1267 &$\pm$&  60 & -0.51 &$\pm$& 0.02& SOAR \\
\object{NGC2249} &  703 &$\pm$&  35 & -0.39 &$\pm$& 0.02& SOAR \\
\object{[SL63]268} & 1125 &$\pm$&  60 & -0.32 &$\pm$& 0.01& Magellan \\
\object{[SL63]353} &  969 &$\pm$&  50 & -0.31 &$\pm$& 0.02& Magellan \\
\enddata
\end{deluxetable}

We present the best-fitting SSP ages and metallicities for the 15 LMC star clusters in Table~\ref{tablmcclstpop}. The uncertainties of age determinations were increased to 5\%\ of the corresponding age values because the statistical errors of 1--2\% do not represent the real quality of measurements. The metallicity uncertainties are provided as they were reported by the spectrum fitting procedure. These metallicities correspond to the average metal enrichment of star clusters \citep{SMVS13}:
\begin{equation}
[\mathrm{Z/H}] = [\mathrm{Fe/H}] + 0.75[\mathrm{\alpha/H}]
\label{eq_zh}
\end{equation}

The sensitivity of the {\sc nbursts} technique to the wavelength coverage as well as systematics and degeneracies of age and metallicity determinations at intermediate and old ages ($t>1$~Gyr) were addressed in detail in \citet{Chilingarian08,Chilingarian09,Chilingarian17}. However, such analysis has never been performed for younger ages in the range of 30 to 1000~Myr, therefore we needed to do some extra tests using the CMD data for young star clusters.

\subsection{Star cluster ages: spectrum fitting vs CMDs}

\begin{figure}
\includegraphics[width=\hsize]{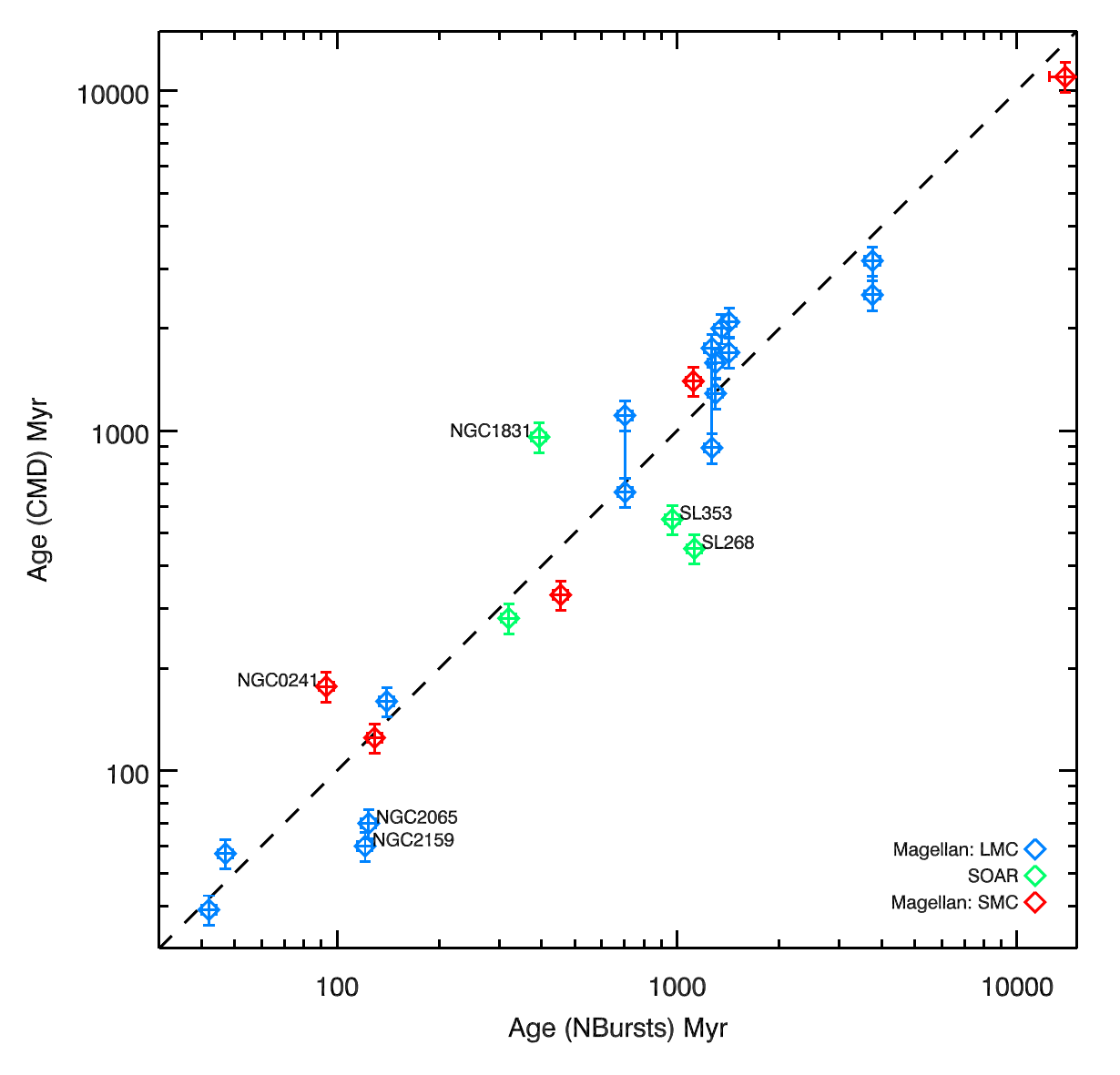}
\caption{Comparison of ages derived from CMDs (literature data) with those determined by the {\sc nbursts} technique. This plot includes measurements for 15 LMC clusters discussed in the paper and for 5 additional SMC clusters with the available Magellan MagE spectra. If multiple CMD measurements were available, they are shown as connected data points. CMD measurements did not have available uncertainties, therefore we assigned a 10\%\ error to them. The 6 strongest outliers are marked. The CMD data sources for the outliers are provided in the text. The CMD measurements for other clusters were taken from \citet{Dirsch2000,2007A&A...462..139K,2008AJ....135.1106G,Niederhofer16,Piatti16,2017MNRAS.468.3150M} \label{fig_age_age}}
\end{figure}

It is important to demonstrate that the full spectrum fitting yields correct age estimates for young and intermediate age star clusters, because stellar population measurements at young ages are subject to the age--metallicity degeneracy similar to old stellar populations \citep{1994ApJS...95..107W} but with a different slope in the age--metallicity space. Therefore, if age estimates are biased, metallicities will also become incorrect.

Fig.~\ref{fig_age_age} shows that the ages determined by the {\sc nbursts} full spectrum fitting technique agree remarkably well with those obtained from CMDs. In this plot we also included our age estimates for 5 Small Magellanic Cloud (SMC) clusters obtained from the analysis of their MagE spectra which will be discussed in a future paper. CMD and spectral ages for 14 out of 20 clusters agree within 10--15\%\ that exceeds our expectations because of known systematic problems of both CMD analysis and full spectrum fitting. 

5 of 6 outliers: \object{NGC~241} in the SMC \citep{1985ApJ...299..211E}, \object{[SL63]268} \citep{vallenari98}, \object{[SL63]353} \citep{Dieball00}, \object{NGC~2065} \citep{Hodge83}, and \object{NGC~2100}) have CMD ages obtained from ground-based seeing limited photometric data, where the source confusion might have biased the results. On the other hand, the NGC~1831 age was estimated by two teams from the HST WFPC2 CMD \citep{2007A&A...462..139K,Niederhofer15} which shows the main sequence turnover at $M_V\sim0.0$~mag that corresponds to $t>700$~Myr, while our MagE spectrum suggests it to be younger than $\sim400$~Myr. It turns out, that both teams assumed zero extinction towards the cluster. However, the visual inspection of Spitzer and AKARI space telescope images reveal that NGC~1831 sits right at the tip of a filament emitting at 8~$\mu$m and in the mid-/far-IR, which suggests that the dust extinction (and maybe even the differential extinction) must play an important role for the CMD analysis of NGC~1831. The color excess $e(B-V)\sim0.15$~mag will shift the CMD up by $\Delta V \approx 0.5$~mag and explain the discrepancy between the photometric and spectroscopic age estimates.

\subsection{Determining $\alpha$-element abundances}

Recent progress in stellar population modeling led to the development of intermediate resolution SSP models with non-solar values of $\alpha$-element abundances. Here we use an extension of MILES v.11 models \citep{SMVS13} computed with differential corrections of MILES stellar spectra using synthetic stellar atmospheres by \citet{Coelho+05} with BaSTI isochrones and the Kroupa IMF. The models employed the same scaling for all $\alpha$-elements and were computed for [$\alpha$/Fe]=$0.0$~dex and [$\alpha$/Fe]=$+0.4$~dex.

Because {\sc nbursts} does not include a native feature to fit an extra parameter of stellar populations in a non-linear minimization loop, in order to derive iron and $\alpha$-element abundances of LMC star clusters we employed an {\it adhoc} approach. We ran {\sc nbursts} 5 times for every spectrum using grids with [$\alpha$/Fe] between 0.0 and $+0.4$~dex with a step of $+0.1$~dex computed by the linear interpolation of between the two MILES based grids and obtained age and [Z/H] estimates for every grid. Then we used the $\chi^2$ values for 5 the best-fitting solution and fitted a second-order polynomial into them in order to estimate the position of the minimum. We avoided the extrapolation and, hence, obtained the best fitting [$\alpha$/Fe] values between $+0.0$ and $+0.4$~dex for each cluster which corresponded to the minimum of $\chi^2$. The [$\alpha$/Fe] uncertainties were computed from the $\chi^2$ statistics. We then interpolated the age and $[Z/H]$ estimates to the best-fitting [$\alpha$/Fe] values.
Finally, [Fe/H] and [$\alpha$/H] are estimated using Eq.~\ref{eq_zh} as: 
\begin{align}
[\mathrm{Fe/H}] = [\mathrm{Z/H}] - 0.75[\mathrm{\alpha/H}] \nonumber \\
[\mathrm{\alpha/H}] = [\mathrm{Z/H}] + 0.25[\mathrm{\alpha/H}]
\end{align}

\begin{deluxetable}{lrclrclrcl}
\tabletypesize{\footnotesize}
\tablecolumns{10}
\tablecaption{Ages, mean metal abundances, and [$\alpha$/Fe] abundance ratios of LMC star clusters derived from the full spectrum fitting against $\alpha$-enhanced stellar population models.\label{tablmcclstpop_alpha}}
\tablehead{
\colhead{Object} & \multicolumn{3}{c}{$t$, Myr} &
\multicolumn{3}{c}{[Z/H], dex} &
\multicolumn{3}{c}{[$\alpha$/Fe], dex} }
\startdata
\object{NGC1651} & 1672 &$\pm$& 118 & -0.53 &$\pm$& 0.02 & 0.26 &$\pm$& 0.03 \\
\object{NGC1831} & 436 &$\pm$& 22 & -0.32 &$\pm$& 0.03 & 0.09 &$\pm$& 0.06 \\
\object{NGC1850} & 46 &$\pm$& 3 & -0.14 &$\pm$& 0.01 & 0.40 &$\pm$& 0.03 \\
\object{NGC1856} & 350 &$\pm$& 18 & -0.33 &$\pm$& 0.01 & 0.18 &$\pm$& 0.03 \\
\object{NGC1863} & 47 &$\pm$& 3 & -0.25 &$\pm$& 0.01 & 0.16 &$\pm$& 0.04 \\
\object{NGC2031} & 171 &$\pm$& 9 & -0.25 &$\pm$& 0.02 & 0.25 &$\pm$& 0.03 \\
\object{NGC2065} & 159 &$\pm$& 8 & -0.32 &$\pm$& 0.03 & 0.25 &$\pm$& 0.04 \\
\object{NGC2155} & 2737 &$\pm$& 239 & -0.52 &$\pm$& 0.03 & 0.15 &$\pm$& 0.03 \\
\object{NGC2159} & 164 &$\pm$& 9 & -0.35 &$\pm$& 0.05 & 0.40 &$\pm$& 0.18 \\
\object{NGC2162} & 1696 &$\pm$& 85 & -0.62 &$\pm$& 0.03 & 0.06 &$\pm$& 0.04 \\
\object{NGC2173} & 2034 &$\pm$& 102 & -0.44 &$\pm$& 0.02 & 0.13 &$\pm$& 0.02 \\
\object{NGC2213} & 1512 &$\pm$& 76 & -0.63 &$\pm$& 0.01 & 0.10 &$\pm$& 0.03 \\
\object{NGC2249} & 652 &$\pm$& 33 & -0.40 &$\pm$& 0.03 & 0.06 &$\pm$& 0.09 \\
\object{[SL63]268} & 1230 &$\pm$& 62 & -0.51 &$\pm$& 0.02 & 0.09 &$\pm$& 0.02 \\
\object{[SL63]353} & 1154 &$\pm$& 58 & -0.47 &$\pm$& 0.03 & 0.00 &$\pm$& 0.04\\ 
\enddata
\end{deluxetable}

In Table~\ref{tablmcclstpop_alpha} we present ages, [Z/H] average metallicities, and [$\alpha$/Fe] abundance ratios for 15 LMC star clusters obtained using the full spectrum fitting as described above.

This is the first occasion when {\sc nbursts} is used to determine $\alpha$-element abundances of stellar populations in a quasi self-consistent way. The detailed study of the stability of this approach, its sensitivity and potential biases lays beyond the scope of this study and will be presented in one of the future papers. Here we provide a brief consistency check with the data obtained using the solar-scaled model grid. The results in this section and their discussion presented below should be considered as a proof-of-concept rather than a final conclusion, which requires a larger number of star clusters to be analyzed.

The age estimates in the case of $\alpha$/Fe fit increase on average by 15\%\ or 0.06~dex with a dispersion of 0.07~dex. Metallicities, on the other hand, become lower by 0.13~dex with a dispersion of 0.10~dex. This is likely caused by the age--metallicity degeneracy that affects the results of the full spectrum fitting \citep[see e.g.][]{Chilingarian08,Chilingarian09}. We should also notice, that because of the way $\alpha$-enhanced MILES based models were constructed using synthetic stellar atmospheres with the highest effective temperature of 7,500$K$, the reliability of the determination of $\alpha$-element abundances at stellar populations younger than $\sim$300~Myr is much lower than that in older star cluster.

\section{Discussion}
\subsection{Age--metallicity relation of the LMC: star clusters vs average stellar population}

\begin{figure}
\includegraphics[width=\hsize]{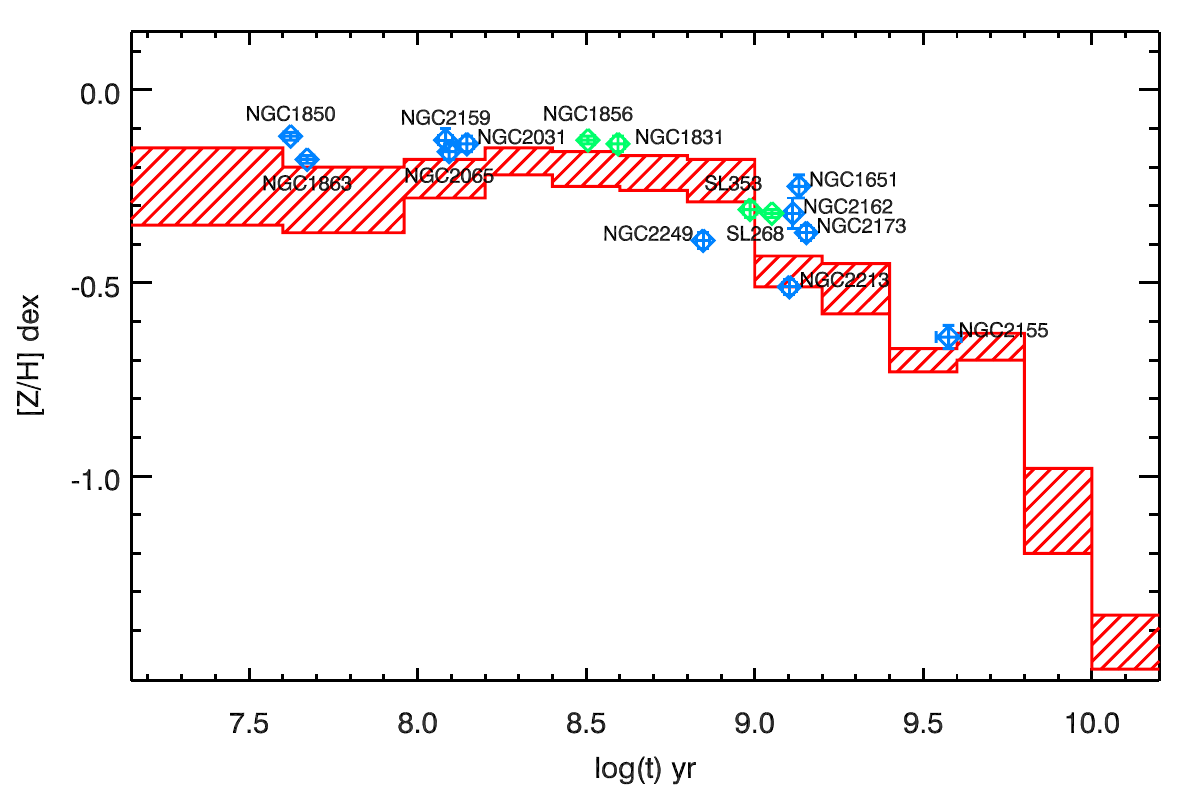}
\caption{Comparison of the age--metallicity diagram for 15 LMC clusters from this study to the CMD based age metallicity relation from \citet{Rubele12} in their tile 8\_3 shown as red shaded areas corresponding to the uncertainties of the metallicity estimates in several age bins.\label{fig_age_met}}
\end{figure}

Our measurements of star cluster stellar population parameters allowed us to build the age--metallicity relation (AMR) for LMC clusters. Now we can compare it to available data published in the literature. In this subsection we refer to the stellar population parameters obtained from the fitting of LMC star cluster spectra against the Solar-scaled grid of MILES based SSP models presented in Table~\ref{tablmcclstpop}.

\citet{Rubele12} presented a new generation of the resolved stellar populations analysis for the LMC based on near-infrared $YHK_s$ CMDs obtained with the ESO VISTA telescope. This color combination is metallicity sensitive, which allowed the authors to recover not only the SFH but also the metal enrichment history in 4 tiles in the LMC covering the area of $\sim$1.4 deg$^{2}$ each. If we exclude their tile 6\_6 that includes the Tarantula nebula and a giant star forming complex making it difficult to measure faint stars, all three remaining tiles demonstrate qualitatively similar age--metallicity relations (see figs.~14--16 in \citealp{Rubele12}). 

Two of the three tiles show that the metallicity decreases by $\sim0.15-0.25$~dex at young ages ($<300$~Myr), which we find very hard to explain from the astrophysical point of view in the framework of the LMC chemical evolution, unless there was a massive supply of metal-poor gas from outside, which we do not see any trace of in terms of an SFR increase at that time. On the other hand, their tile 8\_3 (coordinates ($J2000$): $\alpha=$05:04:53.9, $\delta$=$-$66:15:29) has a nearly flat metallicity distribution for young stars in agreement to what is expected from Eq.~\ref{eq_zevo_solved} provided a very moderate SFR in the last 500--1000~Myr. 

In Fig.~\ref{fig_age_met} we overplot our LMC star cluster AMR on top of the data for the tile 8\_3 from \citet{Rubele12} shown as shaded red boxes with sizes corresponding to their resolution in age and metallicity uncertainties. The two datasets stay in a very good agreement at all ages between 40~Myr (\object{NGC~1865}) and 3.5~Gyr (\object{NGC~2155}). Our metallicity estimates for 5 intermediate age ($0.7<t<2$~Gyr) star clusters are somewhat ($\sim$0.1~dex) higher than the values in the tile 8\_3 and this discrepancy requires more star cluster data to be collected.

\subsection{Reproducing the chemical enrichment history from the observed global SFH}

\begin{figure}
\includegraphics[width=\hsize]{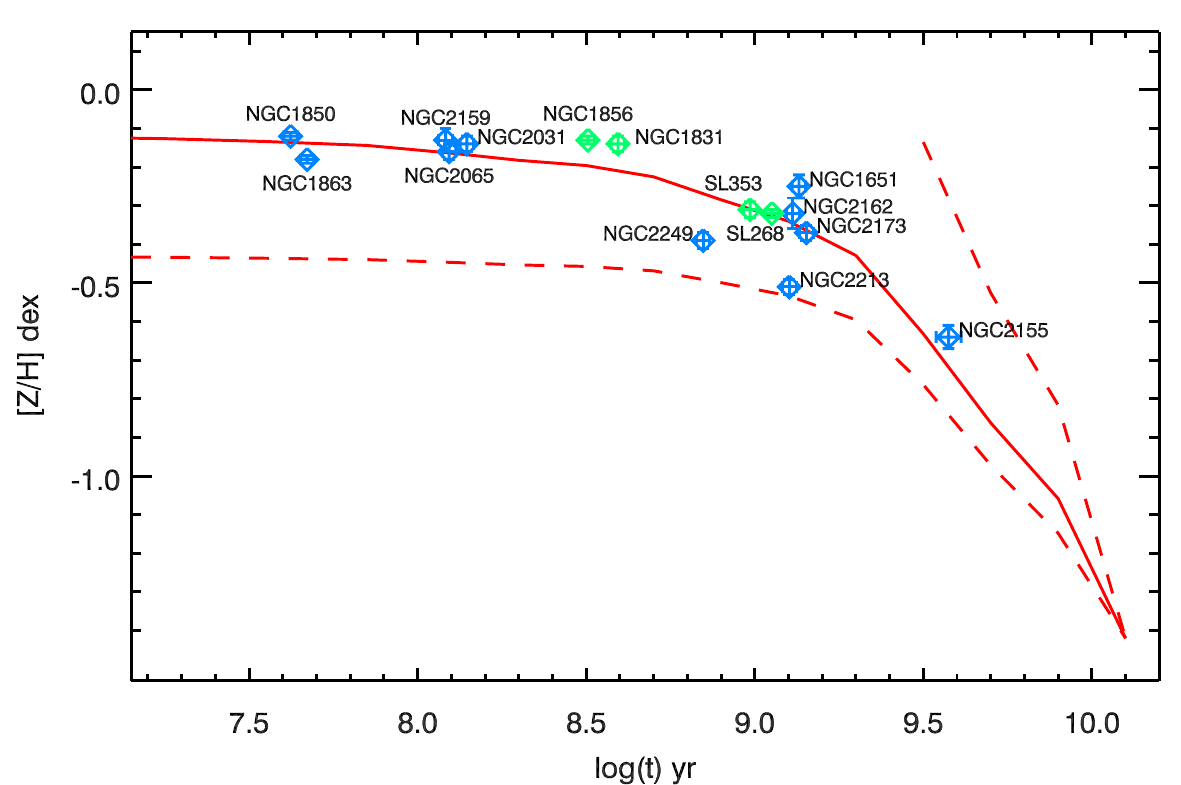}
\caption{Comparison of the age--metallicity diagram for 15 LMC clusters from this study to the predictions of the IRA chemical evolution model based on the LMC global star formation rate from \citep{Harris09} with the best-fitting values of $M_\mathrm{gas}=7.9\cdot10^9 M_{\odot}$ and $\lambda=5.7$ computed for the Chabrier IMF. The solid red line corresponds to the best estimate of the SFR, and dashed lines show the minimal (bottom line) and maximal (top line) total SFR. The dashed line corresponding to the maximal SFR ends at the time when the gas was completely exhausted. \label{fig_age_hz}}
\end{figure}

Now we can compare the AMR of LMC star clusters with the prediction of the chemical enrichment models presented in Section~2. We use the global star formation history of the LMC derived from an optical ground based LMC photometric survey by \citet{Harris09}. 

We accessed the spatially resolved SFH data from \citep{Harris09} using the CDS Vizier service and summed up the SFR estimates in all four metallicity bins across the entire galaxy similar the procedure used by the original authors to produce their fig.~11. We also obtained the upper and lower global SFR limits by using upper and lower SFR limits. Then we computed the chemical evolution models by integrating Eq.~\ref{eq_zevo_solved} for the initial metallicity [Z/H](0)$=-1.42$~dex, which correspond to that of the oldest LMC star clusters excluding several very metal poor ([Fe/H]$<-2$~dex) globular clusters which might have been acquired by the LMC from other Local Group members, and also close to the values reported by \citet{Rubele12} in the oldest age bin. We computed IRA models for a grid of the initial gas mass $M_\mathrm{gas}$ from $2\cdot10^9$ to $1.2\cdot10^{10} M_\odot$ with a step of $10^8 M_\odot$ and a grid of the galactic wind coefficient $\lambda$ from 0 to 9 with a step of 0.1. In parallel, we estimated the evolution of $M_\mathrm{gas}$ in time by integrating Eq.~\ref{eq_mgas} in the IRA framework. We repeated the calculation for $R$ and $y_Z$ values for the Salpeter and Kroupa/Chabrier IMFs. Then we compared the predicted chemical enrichment history to the AMR for LMC star clusters which we presented earlier by interpolating it at the moments of time corresponding to the ages of our star clusters in our sample and then calculating the $\chi^2$ using the statistical uncertainties of star cluster metallicities from the full spectrum fitting procedure. We used the stellar population parameters derived from the fitting of their spectra against olar-scaled SSP models.

For the Salpeter IMF, the minimum of $\chi^2$ with the star cluster AMR is reached at the following values of the parameters of our model: $M_\mathrm{gas}(0)=4.8\cdot10^9 M_\odot$, $\lambda=3.1$. This $\lambda$ value is typical for star-forming galaxies (0.0--5.8) \citep{Spitoni17}. Worth mentioning, that the model by \citet{Spitoni17} included the gas infall as a fundamental assumption under which the $\lambda$ values were calculated. Our IRA model does not include the effect of infall for the reasons stated above, which should lower the expected $\lambda$ value because no external gas supply is happening. Nevertheless, this solution yields the present-day LMC gas mass $M_\mathrm{gas}$(now)=$3.7\cdot10^8 M_\odot$ that is some 25\%\ lower than the total LMC H{\sc i} mass $4.8\pm0.2\cdot10^8 M_\odot$ \citep{2003MNRAS.339...87S} or 30--40\%\ lower than the total mass ($5.8\pm0.5$) estimated assuming the molecular to atomic gas mass ratio of $0.2\pm0.1$ in late type galaxies \citep{1989ApJ...347L..55Y}. Despite that discrepancy, keeping in mind the simplicity of our chemical evolution model, the agreement is good.

The chemical evolution with the Chabrier IMF goes in a slightly different way because of the higher return fraction $R$ and metal yield per generation $y_Z$, therefore in order to fit the observed AMR of star clusters, we need a model with higher values of $M_\mathrm{gas}(0)=7.9\cdot10^9 M_\odot$ and $\lambda=5.7$. This $\lambda$ value is also typical for star-forming galaxies estimated for the Chabrier IMF (0.6--10.2). The present-day gas mass $M_\mathrm{gas}$(now)=$6.2\cdot10^8 M_\odot$ agrees with the observed total gas mass in the LMC within uncertainties. In Fig.~\ref{fig_age_hz} we plot the best-fitting IRA chemical enrichment history model for the Chabrier IMF and the AMR for LMC star clusters.

\begin{figure}
\includegraphics[width=\hsize]{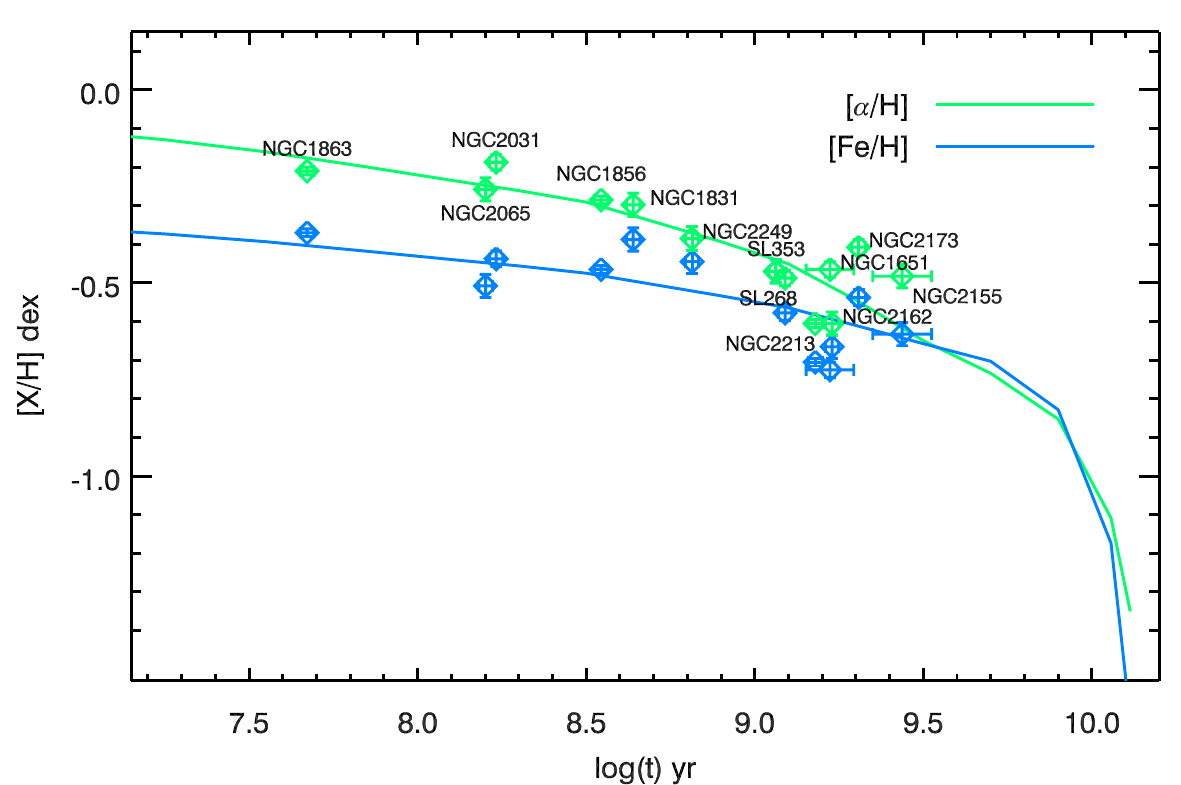}
\caption{Chemical abundances of iron (blue) and $\alpha$-elements (green) derived from the full spectrum fitting of 13 LMC clusters using $\alpha$-enhanced SSP models and the best-fitting models of chemical evolution computed in the IRA+SN~Ia framework for the Kroupa IMF. We excluded NGC~1850 and NGC~2159, where the derived [$\alpha$/Fe] values hit the upper limit of the model grid (+0.4~dex).\label{fig_age_hz_alpha}}
\end{figure}

Finally, we ran the IRS+SN~Ia models for $\alpha$-elements and iron using the same input SFH data from \citet{Harris09} on similar grids of $\lambda$ and $M_\mathrm{gas}(0)$ and assuming the initial abundances corresponding to [Z/H]$_0=-1.42$~dex and [$\alpha$/Fe]$_0=+0.28$~dex. In a similar fashion to the approach described above for the IRA models, we computed $\chi^2$ statistics but this time considering [$\alpha$/H] and [Fe/H] as independent measurements for every LMC cluster. We excluded the two objects from the sample, NGC~1850 and NGC~2159, where the derived [$\alpha$/Fe] values hit the upper limit covered by the models (+0.4~dex). Only one cluster, [SL63]353 has the Solar value of $\alpha$-enhancement and we kept it in the sample.

In Fig.~\ref{fig_age_hz_alpha} we see a satisfactory agreement between the model predictions and observed values of chemical abundances. Because of the recent active star formation in the LMC, we observe $\alpha$-enhanced young and intermediate age (40--400~Myr) star clusters. The best-fitting values for the free parameters, $M_\mathrm{gas}(0) = 1.01\cdot 10^{10} M_\odot$ and $\lambda=6.0$ yields the underestimated total present-day gas mass of about $8\cdot 10^7 M_\odot$. This might be caused by the biases of [Z/H] and/or [$\alpha$/Fe] estimates in the young and intermediate age regime, which is expected to improve when new generations of $\alpha$-enhanced stellar population models become available.

Using star clusters as tracers of chemical evolution of individual elements has an interesting advantage over studies of chemical abundances of individual elements in Milky Way stars in the Solar neighborhood by means of high-resolution spectroscopy. Star clusters allow easy and reliable age estimates while individual stars outside clusters are much harder to date. Therefore, when dealing with star clusters, one can use direct model predictions of [X/H]$(t)$, while for individual stars, in order to remove the age determination uncertainty, it is generally required to operate in the [X/Fe] vs [Fe/H] parameter space \citep[see e.g.][]{MR01,MSRV09,VMS17}. As one can see in Fig.~\ref{fig_age_hz_alpha}, [$\alpha$/H] and [Fe/H] curves cross twice in the case of LMC: the evolution starts from [$\alpha$/Fe]=+0.28~dex, then the enhancement decreases to [$\alpha$/Fe]=-0.05~dex around 6~Gyr ago, then it again starts to grow and reaches [$\alpha$/Fe]=+0.25~dex for young and intermediate age star clusters. Therefore, in case of galaxies with more complex evolutionary histories (e.g. a low-metallicity star burst induced by an accretion of a gas-rich dwarf satellite), the diagnostics in the [X/Fe] vs [Fe/H] parameter space might introduce uncertainties and degeneracies. It may be possible to have several generations of stars of roughly the same [Fe/H] but formed at different epochs and, thus, having different $\alpha$-element abundances. The information on stellar age that is easily accessible in case of star clusters will break this degeneracy. In cases of ``simple'' galaxies like LMC or SMC, a rich dataset that will include several dozens of star clusters covering the entire age range will provide a unique opportunity to study the chemical evolution in great detail and, perhaps, will allow us to distinguish between different progenitors of Type Ia SNe because they will affect the metal enrichment in slightly different ways. Worth mentioning, that the LMC contains only one star cluster in the ``age gap'' between 2.5 and 10~Gyr.

\subsection{Applications of our approach to galaxies beyond the Local Group}

Using a very limited dataset containing only 15 star clusters in the LMC we have demonstrated that our simple approach to study the chemical evolution of galaxies works remarkably well. Not only we reproduced the observed AMR from an observed SFH, but we also obtained the present-day total gas content of the LMC that completely agrees with H{\sc i} observations. With our current 15 clusters we cannot recover the SFH from the AMR directly using Eq.~\ref{eq_psisol} because it is not sampled well enough in age in order to reliably estimate the derivative of the AMR $\dot{Z}(t)$. In the near future We plan to increase our sample of young and intermediate age star clusters in the LMC and then we will be able to directly compare the global SFH derived from star clusters with CMD based measurements.

The analysis of stellar population of star clusters using integrated light spectra can probe star formation histories beyond the immediate vicinity of the Milky Way. A 50,000~$M_\odot$ 10~Gyr old star cluster with the $M_V\approx-5.8$~mag (assuming Kroupa IMF) will have a visual magnitude $m_V=22.0$~mag at a distance of 3.5~Mpc which makes it an easy target for a 6--8~m class telescope in 1--2~h of integration. Younger or more massive clusters will be even brighter: a 1~Gyr old star cluster will have a similar brightness being 10 times less massive. Using multi-object spectroscopy, 100--300 such clusters can be observed in a single exposure, providing an opportunity to study the star formation history of galaxies like \object{Messier~81} or \object{Messier~83} in unprecedented detail. The recent SFH ($t<2$~Gyr) can be characterized using more massive (50,000--100,000~$M_\odot$) clusters out to 10~Mpc with similar investments in the telescope time. No other techniques can be used to recover the SFH with comparable quality at such distances. 

This approach has some caveats. (i) Massive spiral galaxies have star formation and metal enrichment histories that can substantially vary across the galactic disk/bar, e.g. causing radial metallicity gradients. Moreover, the radial migration causes the mixing of stellar population which would translate into spread of the cluster AMR at a given age/radius. Sampling a galaxy cluster population at different radii, however, will allow to estimate variations of the star formation and chemical enrichment histories along the radius. (ii) Mergers with gas-rich dwarf satellites provide supplies of metal-poor gas which would affect the cluster AMR by abruptly decreasing the metallicity of newly formed star clusters -- the detection of this feature in the AMR can provide a strong argument for past merger events, but the chemical evolution equations have to be modified accordingly in order to estimate the star formation history.

\acknowledgments
We thank the anonymous referee for useful suggestions, which helped us to improve the manuscript. I.C.'s research is supported by the Smithsonian Astrophysical Observatory Telescope Data Center. I.C. acknowledges partial support from the M.V.~Lomonosov Moscow State University Program of Development and a Russian--French PICS International Laboratory program (No. 6590) co-funded by the RFBR (project 15-52-15050) for the {\sc nbursts} full spectrum fitting technique development and SOAR data analysis, and from the Russian Science Foundation project 17-72-20119 for the MagE data reduction and analysis and the development of the theoretical models of chemical evolution. R.A.'s research is supported in part by the Mohammed bin Rashid Space Center (MBRSC) grant and FRG17-R-06 Grant from American University of Sharjah. The authors would like to thank A.~Szentgyorgyi and R.~Suleiman for making this collaboration possible. I.C. is grateful to M.~Kurtz, I.~Katkov, O.~Sil'chenko, E.~Grebel, and A.~Vazdekis for useful discussions and suggestions.

\bibliographystyle{aasjournal}
\bibliography{references}

\begin{thebibliography}{}
\expandafter\ifx\csname natexlab\endcsname\relax\def\natexlab#1{#1}\fi
\providecommand{\url}[1]{\href{#1}{#1}}

\bibitem[{{Asa'd} \& {Shahpurwala}(2016)}]{Asad16P}
{Asa'd}, R., \& {Shahpurwala}, A. 2016, \memsai, 87, 676

\bibitem[{{Asa'd}(2014)}]{Asad14}
{Asa'd}, R.~S. 2014, \mnras, 445, 1679

\bibitem[{{Asa'd} \& {Hanson}(2012)}]{Asad12}
{Asa'd}, R.~S., \& {Hanson}, M.~M. 2012, \mnras, 419, 2116

\bibitem[{{Asa'd} {et~al.}(2013){Asa'd}, {Hanson}, \& {Ahumada}}]{Asad13}
{Asa'd}, R.~S., {Hanson}, M.~M., \& {Ahumada}, A.~V. 2013, \pasp, 125, 1304

\bibitem[{{Asa'd} {et~al.}(2016){Asa'd}, {Vazdekis}, \& {Zeinelabdin}}]{Asad16}
{Asa'd}, R.~S., {Vazdekis}, A., \& {Zeinelabdin}, S. 2016, \mnras, 457, 2151

\bibitem[{{Bekki} \& {Chiba}(2005)}]{Bekki05}
{Bekki}, K., \& {Chiba}, M. 2005, \mnras, 356, 680

\bibitem[{{Bertelli} {et~al.}(1994){Bertelli}, {Bressan}, {Chiosi}, {Fagotto},
  \& {Nasi}}]{Bertelli94}
{Bertelli}, G., {Bressan}, A., {Chiosi}, C., {Fagotto}, F., \& {Nasi}, E. 1994,
  \aaps, 106

\bibitem[{{Carrera} {et~al.}(2008){Carrera}, {Gallart}, {Aparicio}, {Costa},
  {M{\'e}ndez}, \& {No{\"e}l}}]{Carrera08}
{Carrera}, R., {Gallart}, C., {Aparicio}, A., {et~al.} 2008, \aj, 136, 1039

\bibitem[{{Chabrier}(2003)}]{Chabrier03}
{Chabrier}, G. 2003, \apjl, 586, L133

\bibitem[{{Chilingarian} {et~al.}(2007{\natexlab{a}}){Chilingarian},
  {Prugniel}, {Sil'Chenko}, \& {Koleva}}]{Chilingarian07}
{Chilingarian}, I., {Prugniel}, P., {Sil'Chenko}, O., \& {Koleva}, M.
  2007{\natexlab{a}}, in IAU Symposium, Vol. 241, Stellar Populations as
  Building Blocks of Galaxies, ed. A.~{Vazdekis} \& R.~{Peletier}, 175--176

\bibitem[{{Chilingarian}(2009)}]{Chilingarian09}
{Chilingarian}, I.~V. 2009, \mnras, 394, 1229

\bibitem[{{Chilingarian} {et~al.}(2008){Chilingarian}, {Cayatte}, {Durret},
  {Adami}, {Balkowski}, {Chemin}, {Lagan{\'a}}, \& {Prugniel}}]{Chilingarian08}
{Chilingarian}, I.~V., {Cayatte}, V., {Durret}, F., {et~al.} 2008, \aap, 486,
  85

\bibitem[{{Chilingarian} {et~al.}(2011){Chilingarian}, {Mieske}, {Hilker}, \&
  {Infante}}]{Chilingarian11}
{Chilingarian}, I.~V., {Mieske}, S., {Hilker}, M., \& {Infante}, L. 2011,
  \mnras, 412, 1627

\bibitem[{{Chilingarian} {et~al.}(2007{\natexlab{b}}){Chilingarian},
  {Prugniel}, {Sil'Chenko}, \& {Afanasiev}}]{Chilingarian07b}
{Chilingarian}, I.~V., {Prugniel}, P., {Sil'Chenko}, O.~K., \& {Afanasiev},
  V.~L. 2007{\natexlab{b}}, \mnras, 376, 1033

\bibitem[{{Chilingarian} {et~al.}(2017){Chilingarian}, {Zolotukhin}, {Katkov},
  {Melchior}, {Rubtsov}, \& {Grishin}}]{Chilingarian17}
{Chilingarian}, I.~V., {Zolotukhin}, I.~Y., {Katkov}, I.~Y., {et~al.} 2017,
  \apjs, 228, 14

\bibitem[{{Chiosi}(1980)}]{Chiosi80}
{Chiosi}, C. 1980, \aap, 83, 206

\bibitem[{{Chiosi} \& {Carraro}(2002)}]{Chiosi02}
{Chiosi}, C., \& {Carraro}, G. 2002, \mnras, 335, 335

\bibitem[{{Clemens} {et~al.}(2004){Clemens}, {Crain}, \&
  {Anderson}}]{2004SPIE.5492..331C}
{Clemens}, J.~C., {Crain}, J.~A., \& {Anderson}, R. 2004, in \procspie, Vol.
  5492, Ground-based Instrumentation for Astronomy, ed. A.~F.~M. {Moorwood} \&
  M.~{Iye}, 331--340

\bibitem[{{Coelho} {et~al.}(2005){Coelho}, {Barbuy}, {Mel{\'e}ndez},
  {Schiavon}, \& {Castilho}}]{Coelho+05}
{Coelho}, P., {Barbuy}, B., {Mel{\'e}ndez}, J., {Schiavon}, R.~P., \&
  {Castilho}, B.~V. 2005, \aap, 443, 735

\bibitem[{{Dekel} \& {Silk}(1986)}]{Dekel86}
{Dekel}, A., \& {Silk}, J. 1986, \apj, 303, 39

\bibitem[{{Dieball} {et~al.}(2000){Dieball}, {Grebel}, \& {Theis}}]{Dieball00}
{Dieball}, A., {Grebel}, E.~K., \& {Theis}, C. 2000, \aap, 358, 144

\bibitem[{{Dirsch} {et~al.}(2000){Dirsch}, {Richtler}, {Gieren}, \&
  {Hilker}}]{Dirsch2000}
{Dirsch}, B., {Richtler}, T., {Gieren}, W.~P., \& {Hilker}, M. 2000, \aap, 360,
  133

\bibitem[{{Elson} \& {Fall}(1985)}]{1985ApJ...299..211E}
{Elson}, R.~A.~W., \& {Fall}, S.~M. 1985, \apj, 299, 211

\bibitem[{{Feast} {et~al.}(2010){Feast}, {Abedigamba}, \&
  {Whitelock}}]{2010MNRAS.408L..76F}
{Feast}, M.~W., {Abedigamba}, O.~P., \& {Whitelock}, P.~A. 2010, \mnras, 408,
  L76

\bibitem[{{Francis} {et~al.}(2012){Francis}, {Drinkwater}, {Chilingarian},
  {Bolt}, \& {Firth}}]{Francis12}
{Francis}, K.~J., {Drinkwater}, M.~J., {Chilingarian}, I.~V., {Bolt}, A.~M., \&
  {Firth}, P. 2012, \mnras, 425, 325

\bibitem[{{Glatt} {et~al.}(2008){Glatt}, {Gallagher}, {Grebel}, {Nota},
  {Sabbi}, {Sirianni}, {Clementini}, {Tosi}, {Harbeck}, {Koch}, \&
  {Cracraft}}]{2008AJ....135.1106G}
{Glatt}, K., {Gallagher}, III, J.~S., {Grebel}, E.~K., {et~al.} 2008, \aj, 135,
  1106

\bibitem[{{Greggio} \& {Renzini}(1983)}]{GR83}
{Greggio}, L., \& {Renzini}, A. 1983, \memsai, 54, 311

\bibitem[{{Harris} \& {Zaritsky}(2009)}]{Harris09}
{Harris}, J., \& {Zaritsky}, D. 2009, \aj, 138, 1243

\bibitem[{{Hodge}(1983)}]{Hodge83}
{Hodge}, P.~W. 1983, \apj, 264, 470

\bibitem[{{Iben} \& {Tutukov}(1984)}]{IT84}
{Iben}, Jr., I., \& {Tutukov}, A.~V. 1984, \apjs, 54, 335

\bibitem[{{Iben} \& {Tutukov}(1985)}]{IT85}
---. 1985, \apjs, 58, 661

\bibitem[{{Iwamoto} {et~al.}(1999){Iwamoto}, {Brachwitz}, {Nomoto},
  {Kishimoto}, {Umeda}, {Hix}, \& {Thielemann}}]{Iwamoto+99}
{Iwamoto}, K., {Brachwitz}, F., {Nomoto}, K., {et~al.} 1999, \apjs, 125, 439

\bibitem[{{Kelson} {et~al.}(2001){Kelson}, {Illingworth}, {Franx}, \& {van
  Dokkum}}]{Kelson01}
{Kelson}, D.~D., {Illingworth}, G.~D., {Franx}, M., \& {van Dokkum}, P.~G.
  2001, \apjl, 552, L17

\bibitem[{{Kenyon} {et~al.}(1993){Kenyon}, {Livio}, {Mikolajewska}, \&
  {Tout}}]{KLMT93}
{Kenyon}, S.~J., {Livio}, M., {Mikolajewska}, J., \& {Tout}, C.~A. 1993, \apjl,
  407, L81

\bibitem[{{Kerber} {et~al.}(2007){Kerber}, {Santiago}, \&
  {Brocato}}]{2007A&A...462..139K}
{Kerber}, L.~O., {Santiago}, B.~X., \& {Brocato}, E. 2007, \aap, 462, 139

\bibitem[{{Kroupa}(2001)}]{Kroupa01}
{Kroupa}, P. 2001, \mnras, 322, 231

\bibitem[{{Livanou} {et~al.}(2013){Livanou}, {Dapergolas}, {Kontizas},
  {Nordstr{\"o}m}, {Kontizas}, {Andersen}, {Dirsch}, \&
  {Karampelas}}]{Livanou13}
{Livanou}, E., {Dapergolas}, A., {Kontizas}, M., {et~al.} 2013, ArXiv e-prints,
  arXiv:1303.2538

\bibitem[{{Mannucci} {et~al.}(2006){Mannucci}, {Della Valle}, \&
  {Panagia}}]{MDVP06}
{Mannucci}, F., {Della Valle}, M., \& {Panagia}, N. 2006, \mnras, 370, 773

\bibitem[{{Marshall} {et~al.}(2008){Marshall}, {Burles}, {Thompson},
  {Shectman}, {Bigelow}, {Burley}, {Birk}, {Estrada}, {Jones}, {Smith},
  {Kowal}, {Castillo}, {Storts}, \& {Ortiz}}]{2008SPIE.7014E..54M}
{Marshall}, J.~L., {Burles}, S., {Thompson}, I.~B., {et~al.} 2008, in
  \procspie, Vol. 7014, Ground-based and Airborne Instrumentation for Astronomy
  II, 701454

\bibitem[{{Martocchia} {et~al.}(2017){Martocchia}, {Bastian}, {Usher},
  {Kozhurina-Platais}, {Niederhofer}, {Cabrera-Ziri}, {Dalessandro},
  {Hollyhead}, {Kacharov}, {Lardo}, {Larsen}, {Mucciarelli}, {Platais},
  {Salaris}, {Cordero}, {Geisler}, {Hilker}, {Li}, \&
  {Mackey}}]{2017MNRAS.468.3150M}
{Martocchia}, S., {Bastian}, N., {Usher}, C., {et~al.} 2017, \mnras, 468, 3150

\bibitem[{{Maschberger} \& {Kroupa}(2011)}]{Maschberger11}
{Maschberger}, T., \& {Kroupa}, P. 2011, \mnras, 411, 1495

\bibitem[{{Matteucci}(1994)}]{Matteucci94}
{Matteucci}, F. 1994, \aap, 288, 57

\bibitem[{{Matteucci}(2012)}]{Matteucci12}
---. 2012, Chemical Evolution of Galaxies, doi:10.1007/978-3-642-22491-1

\bibitem[{{Matteucci}(2016)}]{Matteucci16}
{Matteucci}, F. 2016, in Journal of Physics Conference Series, Vol. 703,
  Journal of Physics Conference Series, 012004

\bibitem[{{Matteucci} \& {Chiosi}(1983)}]{Matteucci83}
{Matteucci}, F., \& {Chiosi}, C. 1983, \aap, 123, 121

\bibitem[{{Matteucci} \& {Greggio}(1986)}]{MG86}
{Matteucci}, F., \& {Greggio}, L. 1986, \aap, 154, 279

\bibitem[{{Matteucci} \& {Recchi}(2001)}]{MR01}
{Matteucci}, F., \& {Recchi}, S. 2001, \apj, 558, 351

\bibitem[{{Matteucci} {et~al.}(2009){Matteucci}, {Spitoni}, {Recchi}, \&
  {Valiante}}]{MSRV09}
{Matteucci}, F., {Spitoni}, E., {Recchi}, S., \& {Valiante}, R. 2009, \aap,
  501, 531

\bibitem[{{Niederhofer} {et~al.}(2016){Niederhofer}, {Bastian},
  {Kozhurina-Platais}, {Hilker}, {de Mink}, {Cabrera-Ziri}, {Li}, \&
  {Ercolano}}]{Niederhofer16}
{Niederhofer}, F., {Bastian}, N., {Kozhurina-Platais}, V., {et~al.} 2016, \aap,
  586, A148

\bibitem[{{Niederhofer} {et~al.}(2015){Niederhofer}, {Hilker}, {Bastian}, \&
  {Silva-Villa}}]{Niederhofer15}
{Niederhofer}, F., {Hilker}, M., {Bastian}, N., \& {Silva-Villa}, E. 2015,
  \aap, 575, A62

\bibitem[{{Nomoto} {et~al.}(2013){Nomoto}, {Kobayashi}, \& {Tominaga}}]{NKT13}
{Nomoto}, K., {Kobayashi}, C., \& {Tominaga}, N. 2013, \araa, 51, 457

\bibitem[{{Nomoto} {et~al.}(1984){Nomoto}, {Thielemann}, \& {Yokoi}}]{NTY84}
{Nomoto}, K., {Thielemann}, F.-K., \& {Yokoi}, K. 1984, \apj, 286, 644

\bibitem[{{Pagel}(1997)}]{Pagel97}
{Pagel}, B.~E.~J. 1997, Nucleosynthesis and Chemical Evolution of Galaxies, 392

\bibitem[{{Piatti} {et~al.}(2017){Piatti}, {Aparicio}, \& {Hidalgo}}]{Piatti17}
{Piatti}, A.~E., {Aparicio}, A., \& {Hidalgo}, S.~L. 2017, \mnras, 469, 1175

\bibitem[{{Piatti} \& {Bastian}(2016)}]{Piatti16}
{Piatti}, A.~E., \& {Bastian}, N. 2016, \mnras, 463, 1632

\bibitem[{{Pietrinferni} {et~al.}(2004){Pietrinferni}, {Cassisi}, {Salaris}, \&
  {Castelli}}]{2004ApJ...612..168P}
{Pietrinferni}, A., {Cassisi}, S., {Salaris}, M., \& {Castelli}, F. 2004, \apj,
  612, 168

\bibitem[{{Rubele} {et~al.}(2012){Rubele}, {Kerber}, {Girardi}, {Cioni},
  {Marigo}, {Zaggia}, {Bekki}, {de Grijs}, {Emerson}, {Groenewegen},
  {Gullieuszik}, {Ivanov}, {Miszalski}, {Oliveira}, {Tatton}, \& {van
  Loon}}]{Rubele12}
{Rubele}, S., {Kerber}, L., {Girardi}, L., {et~al.} 2012, \aap, 537, A106

\bibitem[{{Salpeter}(1955)}]{Salpeter55}
{Salpeter}, E.~E. 1955, \apj, 121, 161

\bibitem[{{S{\'a}nchez-Bl{\'a}zquez} {et~al.}(2006){S{\'a}nchez-Bl{\'a}zquez},
  {Peletier}, {Jim{\'e}nez-Vicente}, {Cardiel}, {Cenarro},
  {Falc{\'o}n-Barroso}, {Gorgas}, {Selam}, \& {Vazdekis}}]{SanchezBlazquez06}
{S{\'a}nchez-Bl{\'a}zquez}, P., {Peletier}, R.~F., {Jim{\'e}nez-Vicente}, J.,
  {et~al.} 2006, \mnras, 371, 703

\bibitem[{{Sansom} {et~al.}(2013){Sansom}, {Milone}, {Vazdekis}, \&
  {S{\'a}nchez-Bl{\'a}zquez}}]{SMVS13}
{Sansom}, A.~E., {Milone}, A.~d.~C., {Vazdekis}, A., \&
  {S{\'a}nchez-Bl{\'a}zquez}, P. 2013, \mnras, 435, 952

\bibitem[{{Schmidt}(1963)}]{Schmidt63}
{Schmidt}, M. 1963, \apj, 137, 758

\bibitem[{{Schroyen} {et~al.}(2013){Schroyen}, {De Rijcke}, {Koleva},
  {Cloet-Osselaer}, \& {Vandenbroucke}}]{Schroyen13}
{Schroyen}, J., {De Rijcke}, S., {Koleva}, M., {Cloet-Osselaer}, A., \&
  {Vandenbroucke}, B. 2013, \mnras, 434, 888

\bibitem[{{Spitoni} {et~al.}(2017){Spitoni}, {Vincenzo}, \&
  {Matteucci}}]{Spitoni17}
{Spitoni}, E., {Vincenzo}, F., \& {Matteucci}, F. 2017, \aap, 599, A6

\bibitem[{{Staveley-Smith} {et~al.}(2003){Staveley-Smith}, {Kim}, {Calabretta},
  {Haynes}, \& {Kesteven}}]{2003MNRAS.339...87S}
{Staveley-Smith}, L., {Kim}, S., {Calabretta}, M.~R., {Haynes}, R.~F., \&
  {Kesteven}, M.~J. 2003, \mnras, 339, 87

\bibitem[{{Tinsley}(1980)}]{Tinsley80}
{Tinsley}, B.~M. 1980, \fcp, 5, 287

\bibitem[{{Vallenari} {et~al.}(1998){Vallenari}, {Bettoni}, \&
  {Chiosi}}]{vallenari98}
{Vallenari}, A., {Bettoni}, D., \& {Chiosi}, C. 1998, \aap, 331, 506

\bibitem[{{Vazdekis} {et~al.}(2010){Vazdekis}, {S{\'a}nchez-Bl{\'a}zquez},
  {Falc{\'o}n-Barroso}, {Cenarro}, {Beasley}, {Cardiel}, {Gorgas}, \&
  {Peletier}}]{Vazdekis10}
{Vazdekis}, A., {S{\'a}nchez-Bl{\'a}zquez}, P., {Falc{\'o}n-Barroso}, J.,
  {et~al.} 2010, \mnras, 404, 1639

\bibitem[{{Vincenzo} {et~al.}(2016){Vincenzo}, {Matteucci}, {Belfiore}, \&
  {Maiolino}}]{Vincenzo16}
{Vincenzo}, F., {Matteucci}, F., {Belfiore}, F., \& {Maiolino}, R. 2016,
  \mnras, 455, 4183

\bibitem[{{Vincenzo} {et~al.}(2017){Vincenzo}, {Matteucci}, \&
  {Spitoni}}]{VMS17}
{Vincenzo}, F., {Matteucci}, F., \& {Spitoni}, E. 2017, \mnras, 466, 2939

\bibitem[{{Whelan} \& {Iben}(1973)}]{WI73}
{Whelan}, J., \& {Iben}, Jr., I. 1973, \apj, 186, 1007

\bibitem[{{Woosley} \& {Weaver}(1995)}]{WW95}
{Woosley}, S.~E., \& {Weaver}, T.~A. 1995, \apjs, 101, 181

\bibitem[{{Worthey}(1994)}]{1994ApJS...95..107W}
{Worthey}, G. 1994, \apjs, 95, 107

\bibitem[{{Young} \& {Knezek}(1989)}]{1989ApJ...347L..55Y}
{Young}, J.~S., \& {Knezek}, P.~M. 1989, \apjl, 347, L55

\end{thebibliography}

\end{document}